\documentclass[10pt,twoside,BCOR7mm,DIV15,headinclude,footexclude,
               cleardoubleempty,idxtotoc]
{scrartcl}

\usepackage{natbib}
\usepackage[font=small,labelfont=bf]{caption}
\usepackage[english]{babel}
\usepackage{graphicx}
\usepackage{hyperref}
\usepackage{scrpage2}
\usepackage{ifthen}
\usepackage{booktabs}
\usepackage{amsmath}
\usepackage{amssymb}
\usepackage{multicol}
\usepackage{float}
\usepackage{hyperref}

\hypersetup{breaklinks=true
,colorlinks=true,linkcolor=black,urlcolor=blue
,citecolor=black}

\addto\captionsenglish{%
}

\makeatletter
\renewcommand{\@biblabel}[1]{}
\renewcommand{\@cite}[2]{%
{#1\ifthenelse{\boolean{@tempswa}}{,#2}{}}}
\makeatother
\ofoot{\thepage}
\ifoot{}

\setcapindent{0em}
\setheadsepline{1pt}

\setkomafont{pagehead}{\normalfont\sffamily}
\setkomafont{pagenumber}{\normalfont\rmfamily}

\makeatletter
\newcommand{\listofcontributions}{\@starttoc{con}}

\newcommand{\l@contribution} {\@dottedtocline{1}{1.5em}{2.3em}}
\makeatother

\newenvironment{contribution}{
\setcounter{section}{0}
\setcounter{figure}{0}
\setcounter{table}{0}
}{
\newpage
\lehead{}
\rohead{}
}

\begin{document}

\setlength{\baselineskip}{2.5ex}

\begin{contribution}
% EXAMPLE AND TEMPLATE FILE FOR PROCEEDINGS OF THE WOLF-RAYET WORKSHOP.
% PLEASE REPLACE THE TEMPLATE TEXT BY YOUR OWN ARTICLE.
% NOTE THAT YOU MUST NOT PROCESS THIS FILE, BUT THE MASTER FILE:
% latex masterfile; dvips masterfile

% RUNNING AUTHOR: PUT AUTHOR NAMED HERE
\lehead{D.\ Vanbeveren \& N.\ Mennekens}

% RUNNING TITLE; SHORTEN THE TITLE IF NECESSARY
% IN CASE OF A ONE-PAGE CONTRIBUTION (POSTER),
% SQUEEZE AUTHORS AND TITLE IN THIS LINE (Author: Title ...)
\rohead{Massive star population synthesis with binaries}

\begin{center}
% FULL TITLE HEADING
{\LARGE \bf Massive star population synthesis with binaries}\\
\medskip

% AUTHORS LIST
{\it\bf D. Vanbeveren$^{1,2}$,  \& N.\ Mennekens$^1$}\\

% AFFILIATIONS
{\it $^1$Vrije Universiteit Brussel, Brussels, Belgium}\\
{\it $^2$K.U. Leuven, Louvain, Belgium}

% ABSTRACT
\begin{abstract}
We first give a short historical overview with some key facts of massive star population synthesis with binaries. We then discuss binary population codes and focus on two ingredients which are important for massive star population synthesis and which may be different in different codes. Population simulations with binaries is the third part where we consider the initial massive binary frequency, the RSG/WR and WC/WN and SNII/SNIbc number ratio's, the probable initial rotational velocity distribution of massive stars.
\end{abstract}
\end{center}

% TEXT OF THE PAPER, TWO-COLUMN STYLE
\begin{multicols}{2}

\section{Introduction}

It is generally accepted that most of the Wolf-Rayet stars are massive hydrogen deficient core helium burning stars. Some WNL stars (using the original abbreviation-definition of Vanbeveren and Conti, 1980) may be core hydrogen burning objects. They are not considered in the present paper. In the late sixties and seventies Roche lobe overflow (RLOF) in binaries was considered as a most plausible process capable to remove the hydrogen rich layers of a star. Furthermore, at the massive star conference in 1971 in Buenos Aires, Kuhi (1973) presented statistical arguments to conclude that all galactic WR stars may be binary components. At the same time large observational data sets became available of X-ray binaries and together with the WR argument it made flourish massive close binary evolution. Some protagonists are B. Paczynski, E. van den Heuvel, I. Iben, A. Tutukov, L. Yungelson, C. De Loore and the Brussels gang. Interested readers may find many massive binary evolution studies published in this period.

The early seventies are also characterised by important breakthroughs in the study of stellar winds of massive stars (e.g., Castor et al., 1975) and the question was raised whether or not WR stars could form via massive single star evolution, single stars that lose their hydrogen rich envelope by stellar winds (the Conti scenario, Conti, 1976). Chiosi et al. (1978) were among the first to demonstrate that this is indeed possible, but it still had to be shown that WR single stars exist (remember Kuhi, 1971). Vanbeveren and Conti (1980) reconsidered the galactic census of WR binaries. They concluded that the Kuhi statistics is biased and that the real galactic WR+OB frequency is no more than 30-40\% a percentage that still holds today.

Since 1980 large evolutionary data sets of massive single stars and binaries were calculated and with these data sets in combination with a description of the different processes that govern binary evolution (e.g., the effect of the supernova explosion on orbital parameters, the treatment of binary mergers, the treatment of the effect on orbital parameters of mass loss from the system during the RLOF, etc.) it became possible to predict theoretically how a massive star population would look like. It was realized that a comparison with the observed population yields important information in order to understand the physics of massive star/binary evolution. 

Meurs and van den Heuvel (1989) are probably among the first authors to study in some detail the massive star population including binaries. The authors focussed on the number of evolved early type close binaries in the Galaxy, in particular the X-ray binary population. 

Portegies Zwart and Verbunt (1996) introduced the skeleton of what would later on become the population code SeBa used in dynamical N-body simulations. In 1996 they used it to discuss the population of massive binaries with compact companions. 

The investigation of the effects of binaries on the population of O-type and WR-type stars also started by the end of the previous millennium (Dalton and Sarazin, 1995; Vanbeveren, 1995; Vanbeveren et al., 1997; De Donder et al., 1997). A detailed description of the Brussels massive star population code was given in Vanbeveren et al. (1998a = Paper I) (see also Vanbeveren et al., 1998b for an extended review) and it was applied in order to predict the galactic number ratios WR/O, WR+OB/WR, WC/WN, O and WR stars with a compact companion, O and WR runaway stars, O and WR single stars but with a binary origin, O and WR single stars originating from a merged binary, the number of real O and WR single stars, etc. Furthermore, De Donder and Vanbeveren (2004) studied the effects of binaries on the chemical evolution of galaxies and it was therefore necessary in order to extend the code so that it was capable to predict the temporal evolution of the massive star/binary population as a function of metallicity Z. 

Since 1998 other research groups wrote massive star+binary population codes with various degree of sophistication, e.g., Nelemans et al. (2001) and Toonen et al. (2012) substantially updated SeBa, Izzard et al. (2004, Binary$\_$c) simulated the population of core collapse supernovae and gamma-ray bursts, Belczynski et al. (2008, Startrack) focussed on relativistic binaries with an application to future gravitational wave detectors. Eldridge et al. (2008) also presented simulations of massive star populations as a function of Z using an extensive grid of single and binary evolutionary computations.  

Spectral synthesis is a powerful tool in order to investigate extragalactic young massive starburst regions in general, the O and WR spectral features in these starbursts in particular. Starburst99 (Leitherer et al., 1999) is extremely useful for galaxies with active star formation but it has to be kept in mind that it does not account for the presence/evolution of massive binaries. The effect of binaries on the spectral synthesis of young starbursts was a main research topic in Brussels (Van Bever et al., 1999; Van Bever and Vanbeveren, 2000, 2003; Belkus et al., 2003) and repeated by Eldridge and Stanway (2009) who essentially confirmed the Brussels results.

De Mink et al. (2013, 2014) introduced an important parameter in massive star population studies: rotation. They showed that the observed distribution of rotation velocities in O-type stars can be explained entirely by the process of RLOF and the possible spin-up of mass gainers/binary mergers. 

In the line of the work of De Mink et al. a study of the rotation velocities of the O-type companions in WR+O binaries may be most illuminating and may help to answer the question if RLOF and mass transfer happened in WR+O progenitors. Mike Shara, Tony Moffat, Gregor Rauw, Dany Vanbeveren et al. just started an observational project aiming at determining these rotational velocities in as many WR+O binaries as possible. We invite everybody interested in joining the et al..

\section{Massive single star + binary population synthesis codes }

The Brussels population code has been described in detail in the list of papers given in the introduction and very recent updates are discussed in Mennekens and Vanbeveren (2014). Rather than repeating all the basic ingredients (once more) we prefer to highlight two ingredients which may be different in different codes: the initial-final mass relation and the LBV/RSG stellar wind mass loss rate formalism.

\subsection{The initial-final mass relationship}

Some codes (e.g., Nelemans et al., 2001; Izzard et al., 2004; Belczynski et al., 2008; De Mink et al., 2013) use single star interpolation formulae proposed by Hurley et al. (2000)  and binary-evolutionary algorithms described by Tout et al. (1997) and Hurley et al. (2002). Other codes (e.g., Eldridge et al., 2008; the Brussels code) use an extended library of detailed binary evolutionary computations. The latter 2 codes and the codes based on the algorithmic method predict massive star populations that differ mainly in the absolute number of mass gainers of interacting-binaries but the differences are not critical and do not affect overall conclusions made in the papers cited above. A much more severe difference is related to the initial mass-final mass  relation ship. As was discussed by Mennekens and Vanbeveren (2014) for stars with initial mass $\ge$ 20 M$_\odot$ the final masses are significantly larger in Hurley et al. based codes than in the Brussels code, possibly due to different stellar wind mass loss rate formalisms and/or alternative convective core overshooting prescriptions during core hydrogen burning. Note that a similar effect is visible in the intermediate mass range (Toonen et al., 2014). Unfortunately, this difference plays a critical role for the predicted population of binaries with at least one compact companion, even more for systems consisting of two compact stars, and as shown by Mennekens and Vanbeveren (2014) also for the predictions of the detection rates of gravitational wave observatories.

\subsection{The stellar wind mass loss rate formalism during the LBV and the RSG phase}

\textbf{LBV.} The Brussels code adopts the LBV scenario of massive binaries as has been introduced by Vanbeveren (1991). It states that the LBV phase is a common evolutionary phase of the most massive stars and that the LBV mass loss rate suppresses the RLOF/common envelope phase in case Br/case Bc/ case C binaries when the mass loser has a mass higher than $\approx$40 M$_\odot$ (see also Mennekens and Vanbeveren, 2014, for a recent argumentation). However, when this mass loser is a member of a case A binary, the star will lose most of its hydrogen rich layers due to RLOF prior to the LBV phase and it is tempting then to assume that the LBV phase does not happen. A case A binary in this high mass range then follows a more canonical evolutionary scenario where the case A RLOF is followed by case Br RLOF.  Mennekens and Vanbeveren (2014) demonstrated that the way how LBV mass loss is implemented in population codes has an enormous effect on the predicted merger rates of double compact star binaries (primarily double black hole binaries are affected) and thus also on the predicted detection rates of forthcoming advanced LIGO detectors.

Note that the story would be completely different if the LBV phenomenon would appear to be related to massive binary mergers.

\noindent \textbf{RSG.} Since the early days when scientist started to investigate the effects of stellar winds on massive star evolution, very conservative formalisms were used in order to study the effects of the RSG wind. Most common was the formalism proposed by de Jager et al. (1988). However, based on observations of Feast (1992), Bressan (1994) concluded that the mass loss rate during the RSG phase of an LMC 20 M$_\odot$ star could be a factor 10 larger than predicted by the de Jager et al. formalism. Vanbeveren (1995) was among the first to investigate the effect of these larger rates on the evolution of 20 Mo - 25 M$_\odot$ single stars. It was concluded that as a consequence of RSG mass loss a 20-25 M$_\odot$ single star may become a WR star. A more throughout discussion of evolutionary computations of massive single stars with the alternative RSG rates and the effect on the overall WR population was presented in Paper I. Salasnich et al. (1999) also re-investigated the effect of new RSG mass loss rates on the evolution of massive stars and essentially arrived at similar conclusions. Note that these alternative RSG rates also significantly affect the evolution of case C binaries, the RSG scenario as it was described in the Vanbeveren et al. papers which states that RSG mass loss may suppress the RLOF in Case C massive binaries. About 15-20 years after the papers of Bressan and Vanbeveren larger RSG rates were also implemented in the Geneva single star evolutionary code (Ekstrom et al., 2012; Meynet et al., 2015). In the latter paper it was concluded that enhanced mass-loss rates during the RSG phase have little impact on the WR population, contrary to the simulations made in Brussels. The difference between the Geneva and Brussels results is most probably due to the post-RSG mass loss rates used in both codes. To illustrate let us consider a 20 M$_\odot$ star. When due to the larger RSG rates this star has lost about 10 M$_\odot$ during the RSG phase, it leaves the RSG region and starts evolving to the blue part of the HR-diagram. The Geneva code calculates the further evolution by using blue supergiant mass loss rates (Vink-rates) and the star never becomes a WR star. However, although the star still has a rather high hydrogen content in its atmosphere at the moment it leaves the RSG phase (typically X$_{atm}$ $\approx$ 0.5), the models have an internal structure similar to WNL stars. In Brussels we therefore decided to continue the further post-RSG evolution by using typical WNL mass loss rates rather than blue supergiant rates. As a consequence the star loses its remaining hydrogen rich layers, becomes a WNE (also here we use the original nomenclature of Vanbeveren and Conti, 1980) and eventually a WC star. The discussion of post-RSG mass loss rates remains open but at least with the Brussels suggestion it is possible to explain the low luminosity WC stars as observed by Sander et al. (2012). In section 3.2 we will add additional support for higher RSG mass loss rates.

\section{Massive single star + binary population synthesis simulations }

Note that in this section we only consider population simulations where binaries are included.

\subsection{The initial massive binary frequency}

First, it is important to realize that accounting for all the physical processes that determine the evolution of binaries, the massive binary frequency (in a population of stars where star formation has been going on for at least a few million years) is smaller that the binary frequency at birth (on the ZAMS).  In all the population simulations that we published since Paper I it is assumed that the massive binary frequency (binaries with initial period $\le$ 10 yr) at birth f $\ge$ 0.7. This latter value is based on the following argumentation. By studying a sample of 67 bright O-type stars Garmany et al. (1980) concluded that 33\% ($\pm$13\%) are primary of a close binary with mass ratio $>$ 0.2 and period P $\le$ 100 days. As discussed in Paper I a population of O-type stars in a field of continuous star formation consists of real single stars, interacting binaries with periods up to 10 years, mergers looking like singles, post-supernova single O-type mass gainers, etc.. Therefore, to recover the results of Garmany et al. (1980) with a population synthesis simulation we had to start with an initial binary frequency f $\ge$ 0.7. 

As was outlined in the introduction, the observed WR+OB binary frequency (in the Galaxy) seems to be not larger than 30-40\%. But also a population of WR stars in a region where star formation is continuous consists of real WR single stars, WR+OB binaries, WR stars resulting from binary mergers, single WR stars resulting from post-SN single OB-type mass gainers etc. and also here we had to start with a very high initial binary frequency in order to explain the observed WR+OB frequency. 

Recent observations of O-type stars in young clusters seem to confirm a high massive binary fraction (Sana et al., 2008, 2009, 2011, 2012; Rauw et al., 2009) and they propose a value $\ge$ 50\%. A similar exercise as the one made in Paper I (and summarized above) was done by De Mink et al. (2014) but using the observations discussed by H. Sana et al.. Also De Mink et al. concluded that in order to recover the observed $\ge$ 50 \% massive binary frequency of H. Sana and co-workers, one has to start with an initial binary frequency $\ge$ 70\%. 
 
\subsection{the RSG/WR number ratio as function of Z}

RSG winds significantly affect the RSG-timescale of a massive star and depending on the post-RSG mass loss formalism (see section 2.2), they also significantly affect the WR-timescale. The RSG (and post-RSG) mass loss formalism therefore significantly affects the predicted RSG/WR number ratio. Remark that our population code is the only single star + binary code that includes the effects of the alternative RSG mass loss rates as discussed above.  Fig. 1 compares predicted and observed RSG/WR number ratios as function of Z. The observations come from Massey (2003) with updates as reviewed by Massey et al. (2013). The predictions holds for a population with a binary frequency at birth = 70\%. The dashed line is based on the simulation of Eldridge et al. (2008), the full lines are the Brussels predictions (predictions depend on parameters who's values are subject to some uncertainty;  varying the values of these parameters yields a maximum and a minimum RSG/WR number ratio, resp. the upper line and the lower line in Fig. 1). The main difference between the simulation of Eldridge et al. and the Brussels one is the RSG mass loss formalism and we are inclined to conclude that the prediction with the alternative (higher) RSG rates fits the observations better, supporting the conclusion of section 2.2.

%-----------One-column figure -----------------------------------
% Note that only the [H] option is allowed for placing 1-column figures!
\begin{figure}[H]
\begin{center}
\includegraphics[width=\columnwidth]{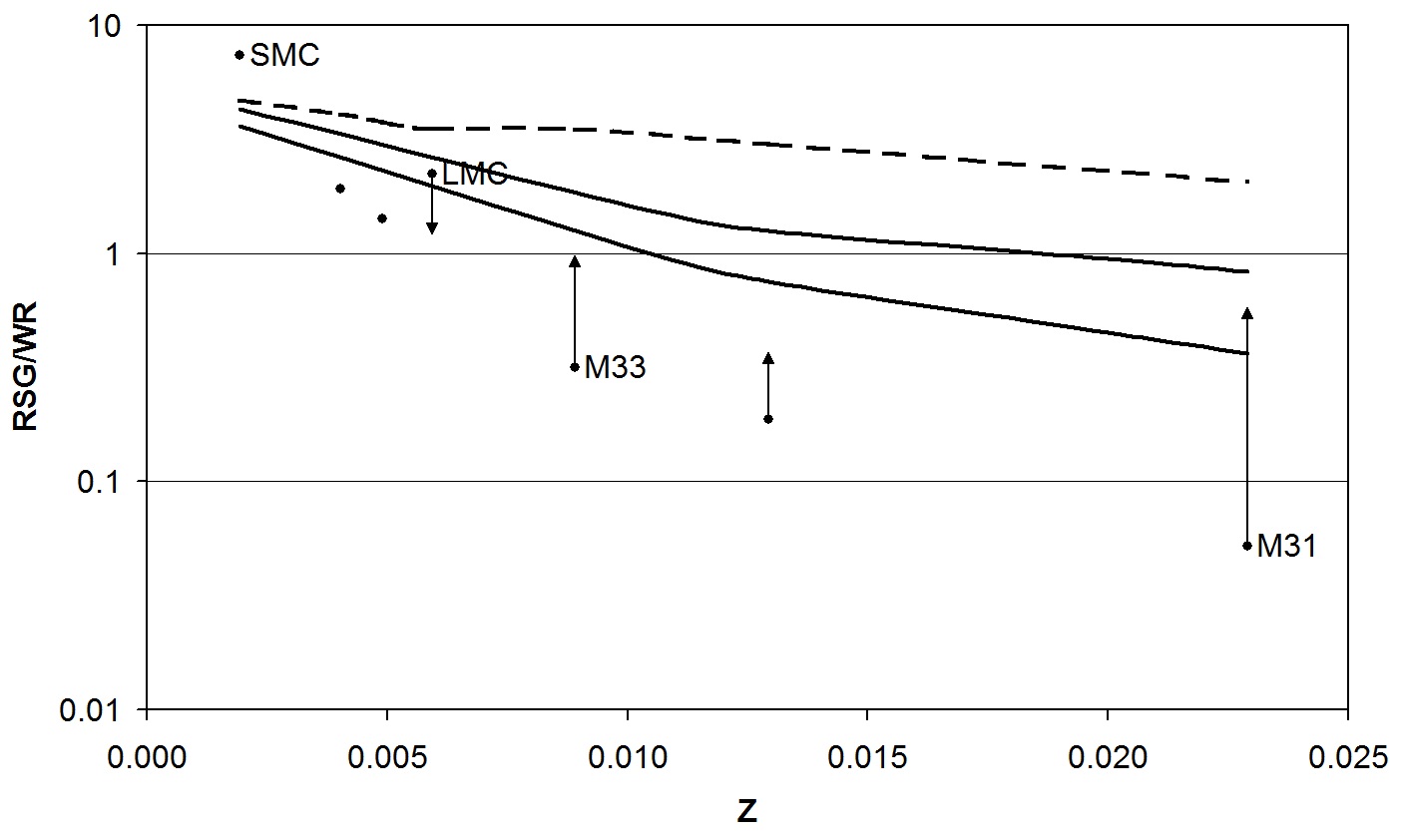}
\caption{A comparison between the observed and predicted RSG/WR number ratio as function of Z. The observations are those reviewed by Massey (2003) and the arrows indicate updates as discussed by Massey et al. (2013). The dashed line is the prediction proposed by Eldridge et al. (2008), the full lines are the maximum/minimum (see text) predictions made with the Brussels code.
\label{example:smallfig}}
\end{center}
\end{figure}
%-----------------------------------------------------------

\subsection{The WC/WN number ratio as function of Z}

L. Smith (1973) argued that metallicity might be responsible for the relative absence of WCs in the Magellanic Clouds, but without understanding the physical mechanism. Vanbeveren and Conti (1980) were among the first to link the effect of Z on stellar wind mass loss and the WC/WN-Z dependence. Detailed population simulations (with binaries) were presented by Vanbeveren et al. (2007) and Eldridge et al. (2008) and compared to observations. It was concluded that correspondence is rather satisfactory. In both studies the observations were those discussed by Massey (2003). It can readily be checked that recent updates (Neugent et al., 2013) do not significantly change the overall conclusions made in the two population studies cited above. 

\subsection{The SNII and SNIbc population}

Studies of the effect of binaries on the population of SNII and SNIbc are numerous and go back to the very beginning of massive binary evolution research. Interested readers may consider Tutukov et al. (1992), Podsiadlowski et al. (1992), Joss et al. (1992), De Donder and Vanbeveren (1998, 2003, 2004), Belczynski et al. (2002) and references therein. Some more recent work essentially confirms the earlier results. One of the conclusions is that most of the progenitors of SN Ibc  are massive binary components with an initial mass $\ge$ 10 M$_\odot$, e.g. most of the progenitors of SN Ibc  do not have an initial mass  $\ge$ 25-30 M$_\odot$ and thus most of the progenitors of SN Ibc are not WR stars. Moreover, since the population of SN II and SN Ibc depends so much on the massive interacting binary population, one may wonder whether SN-population differences between different types of galaxies may reflect differences in the population of these massive binaries.

De Donder and Vanbeveren (1998) compared the overall (cosmological) observed SN II and SN Ibc population with population simulations and it was concluded that the overall cosmological massive interacting binary frequency should be about 50\%.

\subsection{The initial rotational velocity distribution}

The observed rotational velocities of O-type stars in the Galaxy has been discussed by Conti and Ebbets (1977), Penny (1996), and in Paper I. An analysis of these observations in terms of massive single star and close binary evolution was presented in Vanbeveren (2009). It was concluded that a majority of massive O-type stars are born as relatively slow rotators with an average $<$ 200 km/s rather than the 300 km/s used by the Geneva team. The rotation velocity distribution proposed in the papers cited above shows that there is a significant group of rapidly-rotating O-type stars but many of the stars in this latter group are runaways with a peculiar space velocity $>$ 30 km/s. This means that many of them do not have a canonical single star history but are the product of binary evolution (binary mergers, spun-up binary mass gainers, mergers due to dynamical interaction in dense clusters). The latter paper then suggested the following: {\it if one asks whether or not rotation is important for stellar evolution, the answer is yes but perhaps mainly in the framework of binary evolution (rapidly-rotating mass gainers and binary mergers) or in the framework of dynamics in dense clusters where stars collide, merge and become rapid rotators.}

Within the VLT-Flames Tarantula survey Ramirez-Agudelo et al. (2013) investigated the rotational velocities of the O-type stars in 30 Dor and obtained a distribution which is very similar as the one proposed for the Galaxy in the papers listed above.

De Mink et al. (2013) implemented rotation and the evolution of rotation in a population code of single stars and of close binaries. They concluded that starting with an initial population of slowly rotating massive stars (average velocity = 100 km/s), the observed rotational velocity distribution (the one discussed in the papers cited above) can be recovered by properly accounting for all processes that affect the rotation in single star and binary evolution. Therefore (in line with the suggestion of Vanbeveren, 2009 cited above) it can not be excluded that most massive O-type stars are born as slow rotators, much slower than the average value adopted by the Geneva team in their standard evolutionary calculations. If this is true then one may be inclined to conclude that  {\it the overall evolution of massive single stars and of most of the binary components prior to the onset of RLOF hardly depends on rotation}.

\section{Conclusions}

We like to end this paper with an advice and an overall conclusion.

\noindent \textbf{Advice} (not only for young scientists): before starting a research topic try to get a literature overview that is as complete as possible and do not forget that also in the previous millenium interesting studies have been published. 

\noindent \textbf{Conclusion}: a theoretical population study of massive stars where binaries are ignored may have an academic value but may be far from reality.

%\bibliographystyle{aa} % style aa.bst
%\bibliography{myarticle}

\end{multicols}

\end{contribution}

%%-------------------------------------------------------

\end{document}